\documentclass[
aps,%
10pt,%
final,%
notitlepage,%
oneside,%
twocolumn,%
nofootinbib,%
superscriptaddress,%
noshowpacs,%
centertags]%
{revtex4}
\usepackage{graphicx}
\begin{document}

\title{Ionized Gas Characteristics  in the Cavities of the Gas and
Dust Disc of the Spiral Galaxy NGC~6946}

\author{\firstname{Yu.~N.}~\surname{Efremov}}
\email[E-mail: ]{efremovn@yandex.ru} \affiliation{Sternberg State
Astronomical Institute, Moscow State University, Moscow, 119992
Russia}

\author{\firstname{V.~L.}~\surname{Afanasiev}}
\affiliation{Special Astrophysical Observatory, Russian
Academy of Sciences, Nizhniy Arkhyz, 357147 Russia}

\author{\firstname{O.~V.}~\surname{Egorov}}
\affiliation{Sternberg State Astronomical Institute, Moscow State
University, Moscow, 119992 Russia}

\accepted{by Astrophysical Bulletin}

\begin{abstract}
The parameters of the ionized gas in NGC~6946  (in the
[NII]~$\lambda\lambda6548, 6583$, H$_\alpha$ and
[SII]~$\lambda\lambda6717, 6731$ lines) are investigated with the
SAO RAS BTA telescope along three positions of the long slit of
the SCORPIO focal reducer, passing through a number of large and
small cavities of the gaseous disc of the galaxy. Most of these cavities
correspond exactly to the cavities in warm dust, visible at $5-8
\mu$m. We found that everywhere in the direction of NGC~6946 the
lines of ionized gas are decomposed into two Gaussians, one of
which shows almost constant [SII]/H$_\alpha$  and [NII]/H$_\alpha$
ratios, as well as an almost constant radial velocity within the
measurement errors (about $-35\div-50$~km/s). This component is in
fact the foreground radiation from the diffuse ionized gas of our
Galaxy, which is not surprising, given the low   ($12^\circ$)
latitude of  NGC 6946; a similar component is also present in the
emission of neutral hydrogen. The analysis of the component of
ionized gas, occurring in NGC~6946, has revealed that it shows
signs of shock excitation in the cavities of the gaseous disc of
the galaxy. This shock excitation is as well typical for the
extraplanar diffuse ionized gas (EDIG), observed in a number of
spiral galaxies at their high \mbox{Z--coordinates.} This can most
likely be explained by low density of the gas in the NGC~6946 disc
(with the usual photoionization) inside the cavities, due to what
we see the spectral features of the EDIG gas of  NGC~6946,
projected onto them. In the absence of separation of ionized gas into two components by
radial velocities, there is an increasing contribution to the
integral line parameters by the EDIG of our Galaxy when the gas
density in NGC~6946 decreases, which explains some strange
results, obtained in the previous studies. The morphology of warm
dust, visible in the infrared range and HI is almost the same
(except for the peripheral parts of the galaxy, where there are no
sources of dust heating). The shock excitation of the
ionized gas is detected also in the smallest holes, well distinguishable
only in the IR images.
\end{abstract}

\maketitle

\section{INTRODUCTION}
\label{sec:intro:Efremov_n_en}

A close (the distance of about
6~Mpc~\cite{ref_1:Efremov_n_en,ref_2:Efremov_n_en}, so that $1'' =
30$~pc) isolated spiral galaxy NGC~6946, visible almost face-on,
stands out by its record number of both the supernovae observed
there (9), and cavities (121) detected in its HI disc with sizes
reaching up to 2~kpc. The largest of the known collections of  HI
cavities is detected through a detailed study of neutral hydrogen
in this galaxy performed at the Westerbork Synthesis Radio
Telescope (WSRT) \cite{ref_1:Efremov_n_en,ref_2:Efremov_n_en}. We
shall hence use the denominations of the cavities, introduced in
these papers. These data make the NGC~6946 the most suitable
object for understanding the nature of the cavities. This requires
detailed studies of the cavities and their environment in
different regions of the spectrum, determining the parameters of
the ionized gas and the characteristics of star clusters, the
studies of supernova remnants, etc. An ongoing debate persists on
the origin of the cavities, it is described in detail \mbox{in
\cite{ref_1:Efremov_n_en, ref_9:Efremov_n_en, ref_16:Efremov_n_en,
ref_17:Efremov_n_en}.} The hypotheses of  cavity formation
include: the effects of supernova explosions and stellar winds
from hot stars, which in our opinion contradicts the observational
data for large \mbox{cavities \cite{ref_15:Efremov_n_en};} a
result of passage of a massive \mbox{cloud
\cite{ref_17:Efremov_n_en}} or a dark mini-halo
\cite{ref_18:Efremov_n_en}. We will come back to this issue in the
conclusion.

Our initial goal was a re-examination of cavity no.\,85, and the
first study of a huge cavity no.\,107. A comparison of
characteristics of gas in these two cavities is of particular
interest, since these cavities  extremely differ in their
morphology and internal content. The first of them, including a
gigantic queer stellar complex comprising  a  supergiant young
\mbox{cluster \cite{ref_3:Efremov_n_en}} was repeatedly
investigated \mbox{spectrally
\cite{ref_4:Efremov_n_en,ref_5:Efremov_n_en,ref_6:Efremov_n_en}.}
One supercluster seems to be enough to blow this (relatively
small, with the diameter of 700~pc) hole in the gaseous disc.

However, the huge supershell no.\,107 reveals no objects, which
could be responsible for its formation. In this paper we for the
first time  study the cavity spectrally. This is one of the few
objects that does indeed deserve the name of the supershell, since
around it a ring of high HI density   and a rim of increased
brightness are observed in the infrared range at wavelengths of
4.5, 5.8 and $8.0 \mu$m (hot dust emission, associated with the
dense gas and PAH molecules).

This cavity  no.\,107 is the largest (with the diameter of
1.8\,kpc) from the regular round holes in the gaseous disc of
NGC~6946. From the south it is surrounded by an arc of HII
regions, indicating probable star formation, which was initiated
by the collision of the expanding gas shell   with the surrounding
gas, or triggered just when the swept up HI supershell  has
reached a sufficient density. The current diameter of the
structure is much greater than the width of the gaseous disc of
the galaxy, there are no sources of pressure there, and if it
continues to expand, it is only due to inertia. However, the sizes
of all discovered in the gaseous disc of NGC~6946 cavities clearly
exceed its effective depth. This is a general rule for other
spiral galaxies, and remains a difficult problem for the
hypotheses, explaining the formation  and swelling of shells by
any central pressure source possible \cite{ref_2:Efremov_n_en}.
The gas breakthrough beyond the boundaries of galactic disc from
the expanding shell (observed at least twice in our Galaxy,
\mbox{(see, e.g., \cite{ref_7:Efremov_n_en})}, should obviously
lead to a halt in the expansion of the cavity in the galactic
disc. In several cases the signs of this halt (the gas fountains),
are apparently observed in NGC~6946 \linebreak (see
\cite{ref_1:Efremov_n_en, ref_2:Efremov_n_en}).

In addition to the above-mentioned cavities no.\,85 and 107, each
of the three slit positions passes through a number of other
cavities. The objective was to study the variations of  radial
velocity and excitation parameters of the ionized gas along the
slit in hope of getting the data to explain the origin of cavities
in the gas and dust disc of NGC~6946 and other galaxies.

We immediately note that the morphology of warm dust, illustrated by
the image obtained in the far IR range ($5-8 \mu$m) by the Spitzer
Space Telescope, reveals an almost exact conformity with the
picture, revealed in neutral hydrogen, which is correct both for
the cavities and for the regions with high density of gas and
dust. Moreover, a high resolution of the infrared image (compared
with the pattern in neutral hydrogen) allows to discern small
holes in the interstellar medium, practically not detected at the
wavelength of 21\,cm.

\section{OBSERVATIONS}
\label{sec:obs:Efremov_n_en}

The observations were carried out by Afanasiev on August 16--17,
2007 at the  \mbox{6-meter}  BTA telescope of the Special
Astrophysical Observatory of Russian Academy of Sciences (SAO
RAS). We used the SCORPIO focal reducer  (see its description
\mbox{in \cite{ref_8:Efremov_n_en}}) and a long ($6'06''$) slit, the scale
along which amounts to \mbox{$1 \mathrm{px} = 0''.357$.}

Table.~\ref{tab_obs:Efremov_n_en} lists position angles and the
cavities numbers located near the centers of the slits.

\begin{table*}
\begin{center}
\caption{\mbox{Log of observations}}
\label{tab_obs:Efremov_n_en}

\begin{tabular}{c|c|c|c|c|c|c}
\hline
Date & Hole & Position angle, deg. & Spectral range, \AA & Dispersion, \AA & $\mathrm{T_{exp}}$, s & Seeing \\
\hline
 16/17.08.2007 & 107 & $-33 (147)$ & 6100--7100 & 0.52 & 1800 & $2''$ \\
 16/17.08.2007 & 107 & $180 (0)$ & 6100--7100 & 0.52 & 1800 & 2 \\
 16/17.08.2007 & 85 & $-34 (146)$ & 6100--7100 & 0.52 & 1200 & 1.5 \\
 \hline
\end{tabular}
\end{center}
\end{table*}

Figure~1 shows the positions of the spectrograph slits with
respect to the HI cavities. The numbers of cavities from
\cite{ref_1:Efremov_n_en,ref_2:Efremov_n_en} are marked, as well as the cavities, usually
coinciding with them, but smaller in size, recently discovered by
\mbox{Bagetakos et al. \cite{ref_9:Efremov_n_en}.} The slit width amounted to
$1''$, the spectral resolution $FWHM$ of about 3.4\,\AA. The
accuracy of the wavelength scale along the lines of the night sky
is $20-30$~km/s for the lines, obtained with the S/N ratio of
about 5 or higher (as can be clearly seen from the error bars in
the figures, where the distribution of radial velocities along the
slit is given).

The spectrum of the night sky (at the coordinates
$\mathrm{RA=00^h10^m00^s.00}$ and $\mathrm{DEC=+18^\circ
00'00''.0}$; the galactic latitude of about $-42^\circ$ and the
distance from the NGC~6946 of about $50^\circ$) was observed
immediately after NGC~6946 under the same conditions as the galaxy
itself. The spectra obtained were used to subtract the night sky
from the spectrograms of the object. The choice of the position
angle of about $-33^\circ$ was made hoping that at this angle
(parallel to the minor axis of the galaxy) the effect of its
rotation on   radial velocity is minimal, and local features of
radial velocities should be visible.

\begin{figure*}

 \includegraphics[width=14cm]{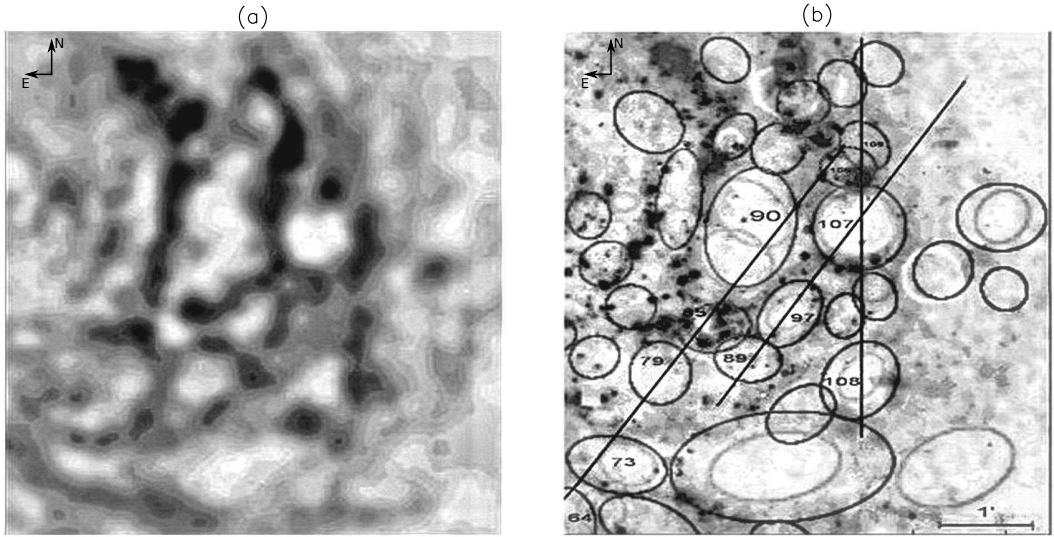}
\caption{The Western region of the NGC~6946 galaxy  in the line of
neutral hydrogen. (a) the image in the HI line according \mbox{to
\cite{ref_1:Efremov_n_en};} (b) the image in the line of HI (in
gray, based on the data from \cite{ref_9:Efremov_n_en}) and the
HII region (in black). We marked the positions of cavities
(according to \cite{ref_1:Efremov_n_en,ref_2:Efremov_n_en,
ref_9:Efremov_n_en}), their numbers according to
\cite{ref_1:Efremov_n_en,ref_2:Efremov_n_en} and the positions of
our slits. The gray and white outlines of the cavities are given
based on \cite{ref_9:Efremov_n_en}.}
\end{figure*}

\begin{figure*}

 \includegraphics[width=15cm]{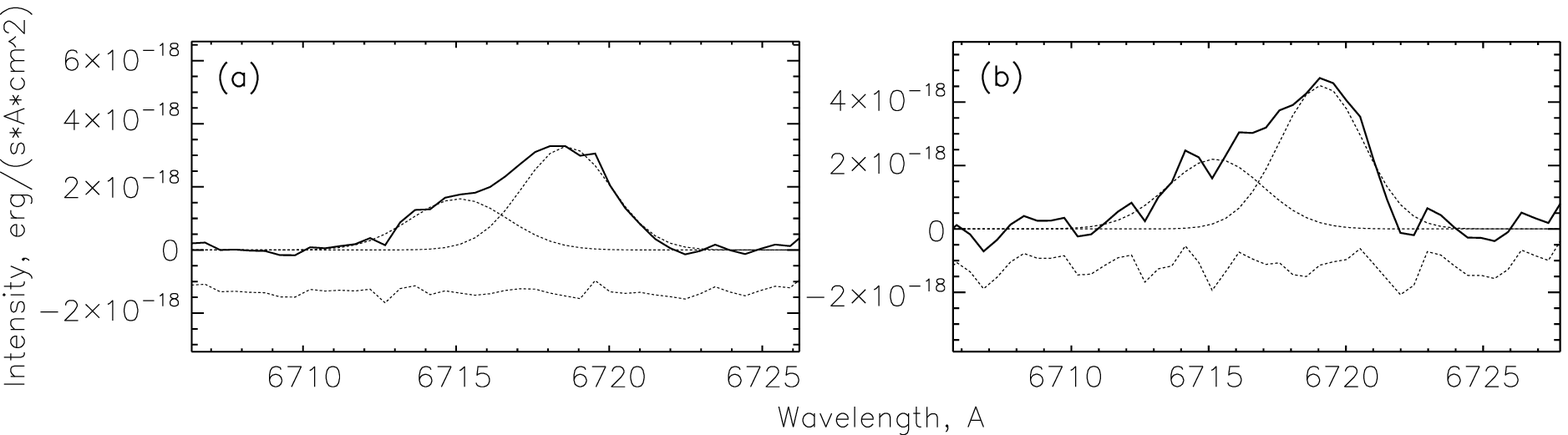}
\caption{An example of decomposition of the line profiles of
sulfur  [SII] $\lambda$6717 into two components. The observed
profile is shown by the solid line, the decomposition into 2
Gaussians and the remainder of their subtraction---by the dashed
line. The right component in the decomposition corresponds to
NGC~6946, the left component---to our Galaxy. The remainder of
subtraction of the Gaussians from the observed spectrum is shifted
down for clarity, the median value of the remainder is zero.}
\end{figure*}

\section{DETECTION OF TWO COMPONENTS IN THE LINE PROFILES OF IONIZED GAS}

In the process of observations  Afanasiev and  Silchenko noticed
that the doublet of forbidden lines of sulfur (they are free from
the sky emission) is observed not only throughout the galaxy, but
also very far beyond its visible boundaries in the spectrograms
obtained higher by declination  $2^\circ$ from NGC~6946.

It prompted an idea that these lines may belong to the foreground,
our Galaxy, which is quite possible, given the relatively low
($11^\circ.7$) latitude of NGC~6946. Therefore,  as shown in
Fig.~2, we performed a Gaussian decomposition of the
radial-velocity profiles of the H$_\alpha$, [NII] and [SII] lines
into two components (bearing in mind that the last two lines are
doublets).

It turned out that along the entire slit there is a component that
has a virtually constant radial velocity over the entire length of
the spectrum (usually in the range of $-40 \pm 7$~km/s) and a
constant within the errors flux ratio in the lines of
H$_\alpha$/[NII]   and   H$_\alpha$/[SII]6717   (Fig.~3). The
velocity obtained from the second component varies along the
spectrum and is closely correlated with the details of morphology
of the gas distribution in NGC~6946. It follows that the first
component originates in the diffuse ionized gas of our Galaxy,
while the second component comes from the NGC~6946. This can be
seen directly in the spectra (including the lines of the night
sky), which show the regions including the [NII] doublet,
H$_\alpha$ and the  [SII] doublet in all the three slits
\mbox{(Figs.~4--6).}

These images show the same smooth and sharp boundary in  the
lowest velocities, especially in the sulfur lines, clean of the
night-sky lines. This boundary is defined by the emission
coming from our Galaxy. It represents the velocity of extraplanar
diffuse ionized gas (EDIG) of the Milky Way in the direction of
NGC~6946. The upper boundary is outlined by the ionized gas of
NGC~6946 itself, with its sharply inhomogeneous intensity
     and variable radial velocities.

\begin{figure*}

 \includegraphics[width=16cm]{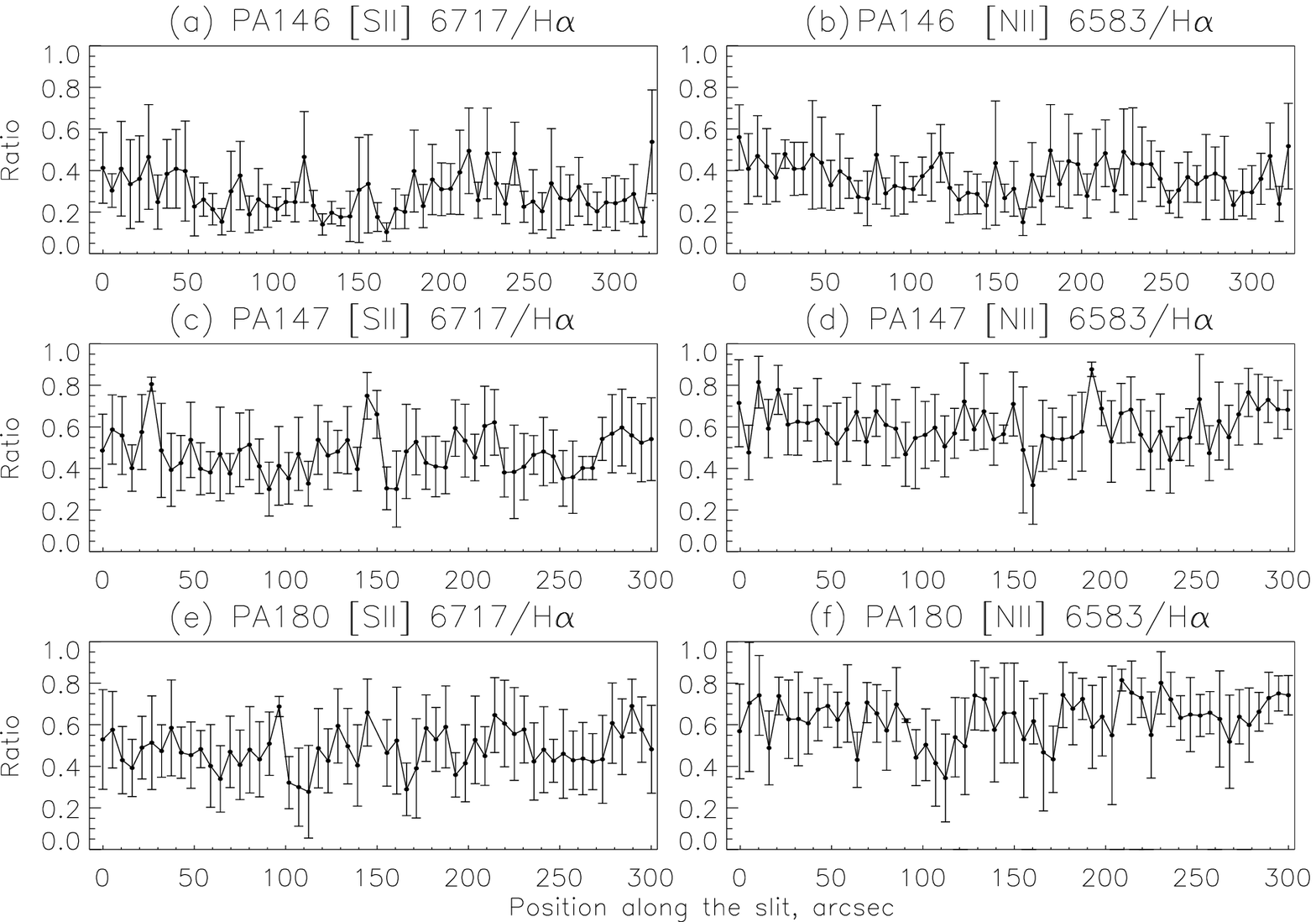}
 \caption{The flux ratio (along all three slits) in the lines of [SII]
$\lambda$6717 and [NII] $\lambda$6583 to the H$_\alpha$  line flux
 in the component of ionized gas, the radial
velocity of which points to the belonging to our Galaxy. The
similarity of line profiles along the slit (especially noticeable
with the slit position PA~147) confirms their origin in the same
discrete warm ionized medium (WIM) clouds of our Galaxy.}
\end{figure*}

\begin{figure*}

\includegraphics[width=15cm]{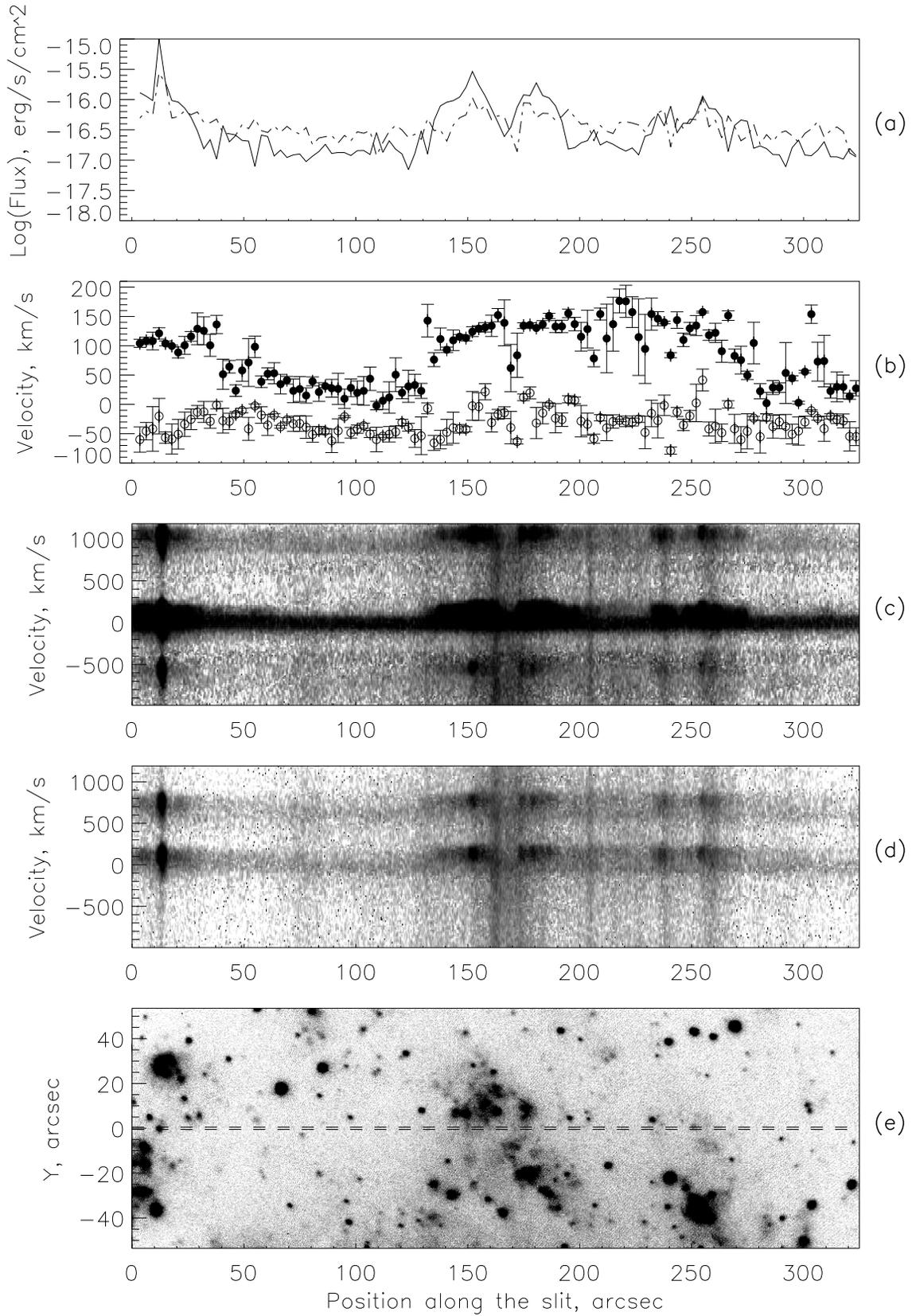}
\caption{The variations of spectral characteristics along the slit
position PA 146, passing through the cavity no.\,85: (a) the
fluxes in H$_\alpha$ from NGC~6946 (the solid line) and the WIM of
our Galaxy (the dashed-dotted line); (b) radial velocities of
ionized gas of NGC~6946 (the upper curve) and the WIM of the
Galaxy (the bottom curve) in the H$_\alpha$ line; (c) a part of
the spectrum in the region of the H$_\alpha$ line. The wavelength
scale is transformed into the velocity scale relative to the
H$_\alpha$ $\lambda$6562.8 line; (d) similar to the plot (c) for
the [SII] lines. The wavelength scale is given relative to the
[SII] $\lambda$6717; (e) the slit position in the H$_\alpha$ image
of the galaxy.}
\end{figure*}

\begin{figure*}

 \includegraphics[width=12cm]{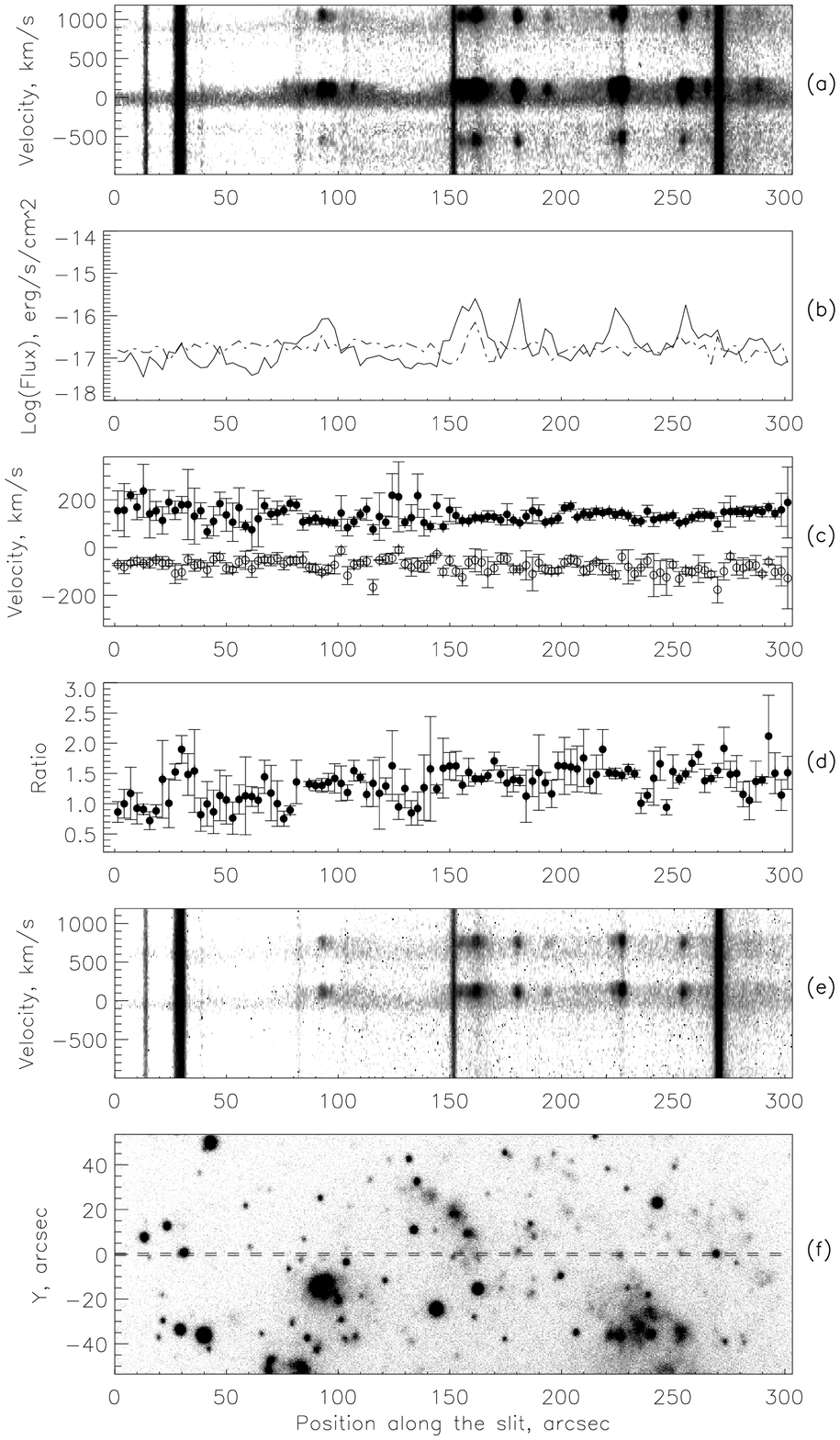}
 \caption{The variations of spectral characteristics along the slit position PA
147, passing through the cavity no.\,107: (a) a part of the
spectrum in the region of the H$_\alpha$ line. The wavelength
scale is converted into the velocity scale  relative to the
H$_\alpha$ $\lambda$6562.8 line; (b) the fluxes in the H$_\alpha$
line from NGC 6946 (the solid line) and the WIM of our Galaxy (the
dash-dotted line); (c) radial velocities of ionized gas of
NGC~6946 (the upper curve) and the WIM of our Galaxy (the bottom
line) in the [SII] $\lambda$6717 line; (d) the flux ratio in
components of the doublet line   [SII]
F($\lambda$6717)/F($\lambda$6731); (e) similar to the plot (a) for
the [SII] lines. The wavelength scale is given relative to the
[SII] $\lambda$6717 line. (f) the slit position in the
H$_\alpha$-image of the galaxy. }
\end{figure*}

The presence of a very substantial admixture of the foreground
ionized gas emission has strangely never been detected in the
previous studies (including ours),  and its indirect signs
(described, for example, in \cite{ref_19:Efremov_n_en} as far back as in 1988)
were attributed to a poor accounting of the background sky, or
else the peculiar radial velocities in several regions of
NGC~6946. The flux from the ionized gas of the foreground is
demonstrated in Figs.\,4--9 by the dashed line, outside the HII
regions it usually exceeds the flux from NGC~6946. Note that
Fig.~2 was built exactly from the spectral regions of NGC~6946,
corresponding to the HII regions.

\section{RADIAL VELOCITIES}

The properties of the foreground gas will be discussed below, for
now we will discuss exclusively the NGC~6946. Figures~7, 8 and 9
demonstrate the radial velocities in
three lines of ionized gas both for NGC~6946 and for the
foreground for the three slit positions, as well as the radial velocity in HI along all the
slit positions, that were kindly provided by R.\,Boomsma (for
details of his observations, see \cite{ref_1:Efremov_n_en}).

The averaging was carried out in the regions sized $ 2.8''$. First
of all, note that the velocity dispersion (or rather, the scatter
of measurements from point to point) sharply increases  in the
cavities. This may be a sign of their expansion, but, more probably, a
consequence of the lower signal-to-noise ratio for the gas flux of
lower density in the cavities.

\begin{figure*}

 \includegraphics[width=15cm]{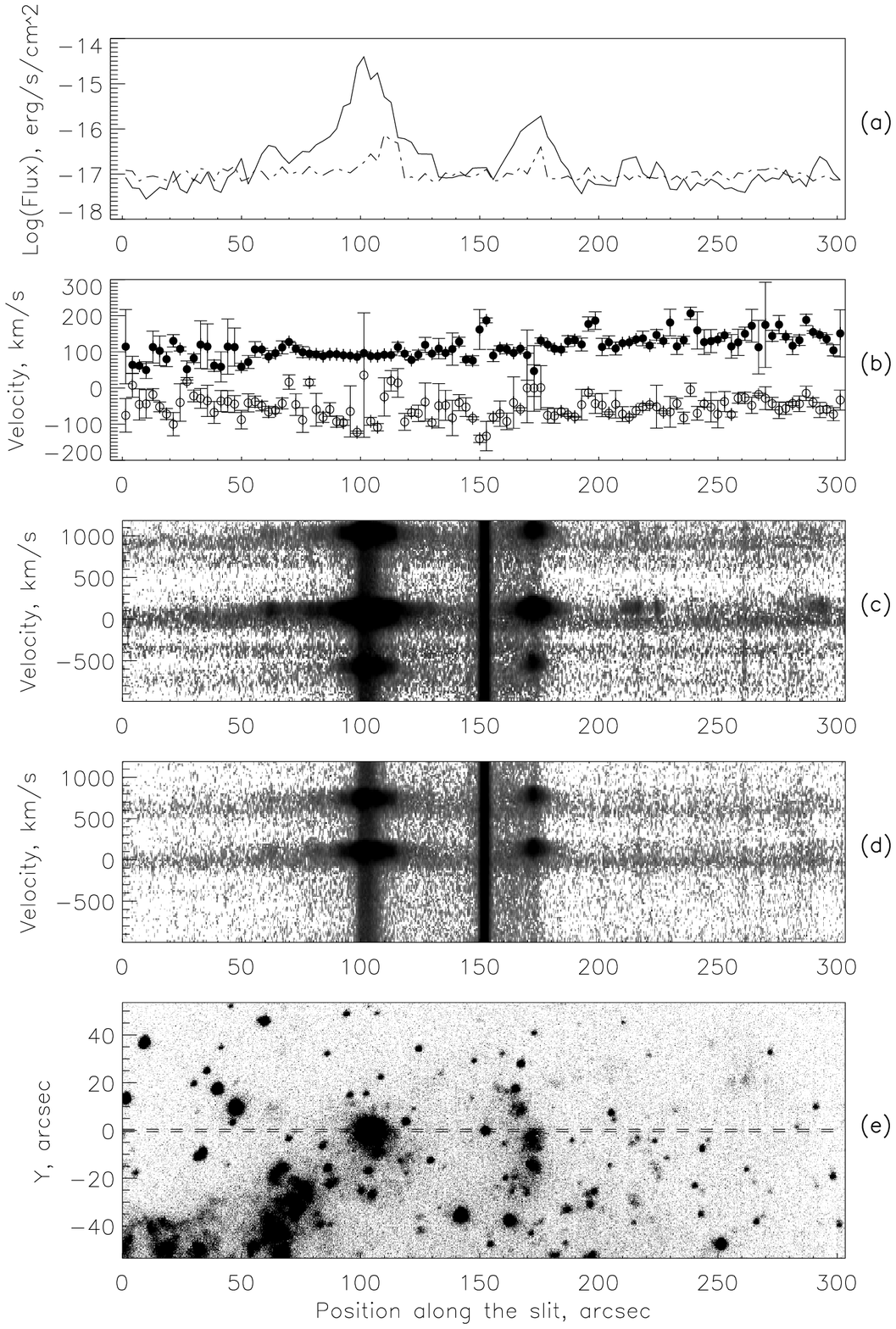}
  \caption{The variations of spectral characteristics along the slit position PA
180, passing through the cavity no.\,107. The plots (a)--(e) are
similar to those shown in Fig.~4.}
\end{figure*}

\begin{figure*}

 \includegraphics[width=15cm]{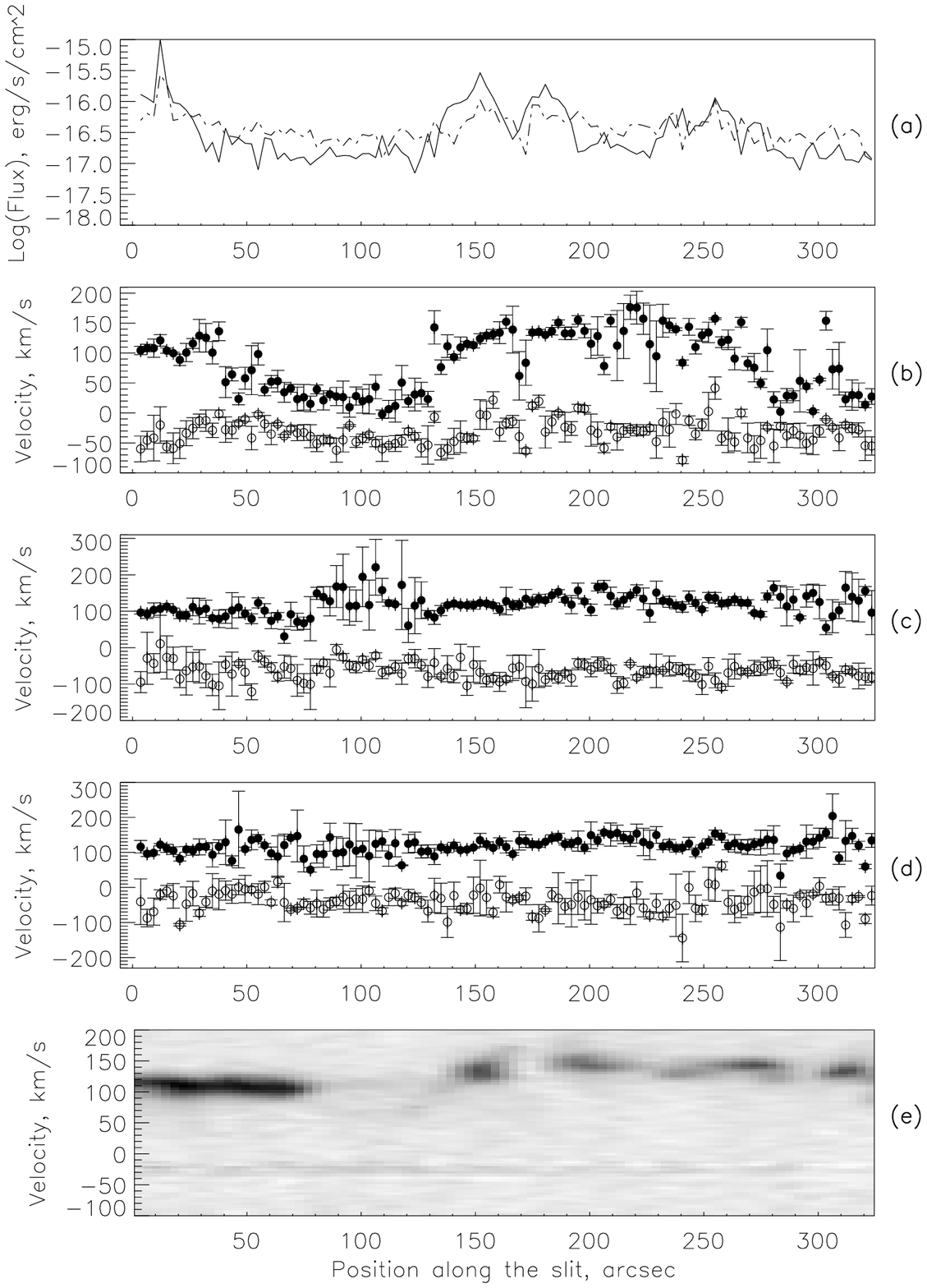}
 \caption{The variations of spectral characteristics along the slit
position PA 146, passing through the cavity no.\,85: (a) the
H$_\alpha$ line fluxes   from NGC~6946 (the solid line) and the
WIM of our Galaxy (the dashed-dotted line); (b) radial velocities
of ionized gas in NGC~6946 (the upper curve) and in the WIM of our
Galaxy (the bottom curve) in H$_\alpha$; (c) the same, in the
[SII] $\lambda$6717 line; (d) the same, in the [NII] $\lambda$6583
line; (e) radial velocities in the HI 21\,cm line.}
\end{figure*}

Sometimes a comparison of radial velocities of different species
of ionized gas reveals intriguing differences. In the cavities
no.\,90 and 73 the radial velocities in H$_\alpha$ drastically
deviate from the mean towards the negative velocities, whereas the
[SII] and [NII] lines  give an almost constant velocity. Moreover, the lines of sulfur reveal a deviation
towards the positive velocities in the cavity no.\,90 (Fig.~7). In this cavity neutral
hydrogen splits into two components based on velocity, one of
which continues to smoothly move along the slit, while the other
indicates that one of the walls of the cavity is approaching us,
deviating in the same direction as the H$_\alpha$ velocities,
while ionized hydrogen shows an even greater velocity of this
wall. Perhaps, this points to the outflow of gas from the shell.

The expansion of the much thicker than the HI disc of the
galaxy cavity seems inexplicable, as already noted \mbox{(see also
\cite{ref_2:Efremov_n_en}).} It is possible that in the case of cavity no.\,90
we are  dealing not with an expanding cavity in the gaseous disc,
but simply with an unusually sharply limited  interarm region of
low density (as already suspected in \cite{ref_1:Efremov_n_en}). However, the
cavity no.\,90 is shown in \cite{ref_9:Efremov_n_en} to be double, what can be
seen in Fig.~1. The diversity in the behavior of radial velocities
of different species of ionized gas   inside  this super-giant
cavity (Fig.~7, the interval along the slit is
\mbox{$80''$--$140''$}) can not be unambiguously explained.

Note also the curious ``peaks'' (the deviations towards higher
velocities) observed in the cavities no.\,90 and no.\,107 solely
in the lines of [SII] and [NII], respectively (Figs.\,7 and 8).
Note also that the lines of [SII] of the foreground give the
velocity peak in the positive direction in the region of
high-density of ionized gas in NGC~6946. This can be explained
either by accidence, or by a not quite accurate for such densities
separation of spectral lines into the components coming from
NGC~6946 and the foreground.

Particularly note in Fig.~7 a narrow dip in the velocity along the
H$_\alpha$ line,  corresponding to a small region of lowest gas
density and the lowest  radial velocity inside the hole no.\,85,
the center of which is located   $ 7''$ to the east of the
supercluster. This region was found in \cite{ref_4:Efremov_n_en} and called a
deep dip in \cite{ref_5:Efremov_n_en}. A hint at this dip can also be found in
the [SII] line (but not in the [NII] line) \mbox{in Fig.~7.}

Even inside the largest cavities (nos.\,90 and 107) we do not
observe any features in the radial velocity field of ionized gas,
similar for all the lines. Therefore, the hopes of finding the
expansion of the cavities have been unfulfilled. The behavior of
radial velocities of neutral hydrogen within the cavities is also
ambiguous; it was discussed by Boomsma \cite{ref_1:Efremov_n_en}.

\section{LINE FLUX RATIOS}

We found that within the cavities the flux ratios [SII]/H$_\alpha$
are systematically higher than beyond their limits, whereas for
the HII regions, including those located at the periphery of the
cavity no.\,107, the flux ratios correspond to those, normal for
the photoionization of gas by hot stars.

A supergiant complex of HII regions  is located at the northern
edge of the second in size and
brightness in NGC~6946 supershell no.\,107. This complex is situated at the junction
of cavities no.\,107 and 106 (Figs.~1 and 10), and could be
regarded as an illustration of the possibility of formation of
giant star-forming regions via collisions of two expanding shells
of gas.

The hypothesis of star formation, induced by the shock wave
collision, was proposed by Chernin \mbox{et al. \cite{ref_10:Efremov_n_en}} and
then repeatedly supported by other authors. This HII complex,
located next to the HII cavity  no.\,107 is crossed by our slit
position PA 180. Its  spectrum did not
reveal any signs of shock excitation, despite our expectations, on the contrary, the ratios
of both [SII]6717+6731 and [NII]6583 to H$_\alpha$ are minimal
here (Fig.~10).

It is considered (see, e.g. Moiseev \mbox{et al. \cite{ref_11:Efremov_n_en}})
that the value of flux ratio in these lines in excess of 0.4
indicates that the main contribution into the gas ionization is
made by the shock waves. In Fig.~11 we drew vertical lines through
the points of maximum values of the [SII]/H$_\alpha$ ratio, and a
horizontal line through the value of this ratio equal to 0.4. We
can see that the slit positions where it exceeds 0.4
correspond to the HI cavities, more accurately---to the boundaries
of the cavities, outlined by warm dust. Two narrow peaks of the
[SII]/H$_\alpha$ ratio (at the coordinates along the slit of
$170''$ and $240''$ on both sides of the cavity no.\,97)
accurately correspond to the positions on the slit occupied by two
mini-cavities in warm dust, almost indistinguishable in the HI
images.

\begin{figure*}

 \includegraphics[width=16cm]{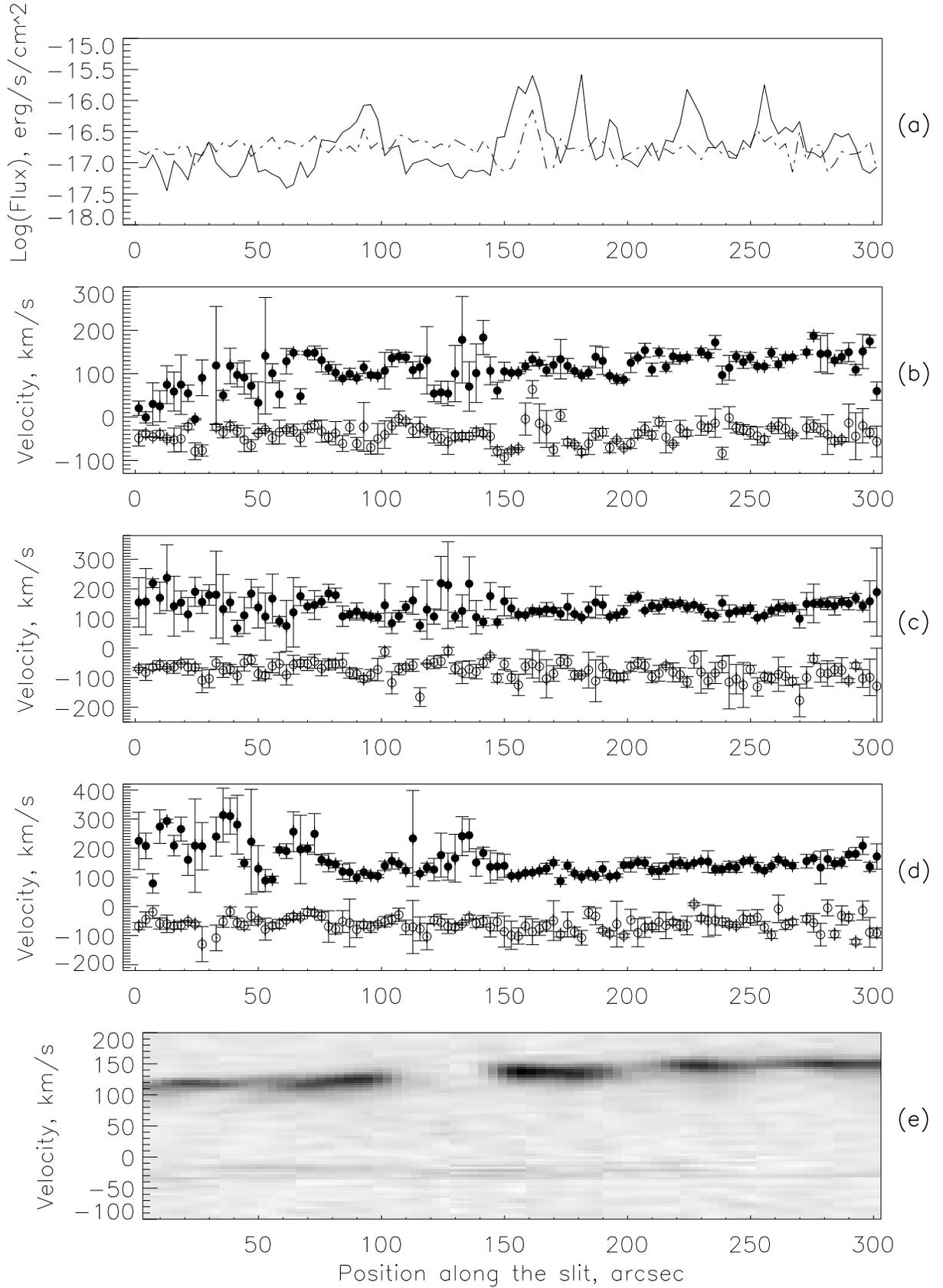}
  \caption{The variations of spectral characteristics along the slit PA
147, passing through the cavity no.\,107. The plots (a)--(e) are
similar to those shown in Fig.\,7.}
\end{figure*}

\begin{figure*}

 \includegraphics[width=16cm]{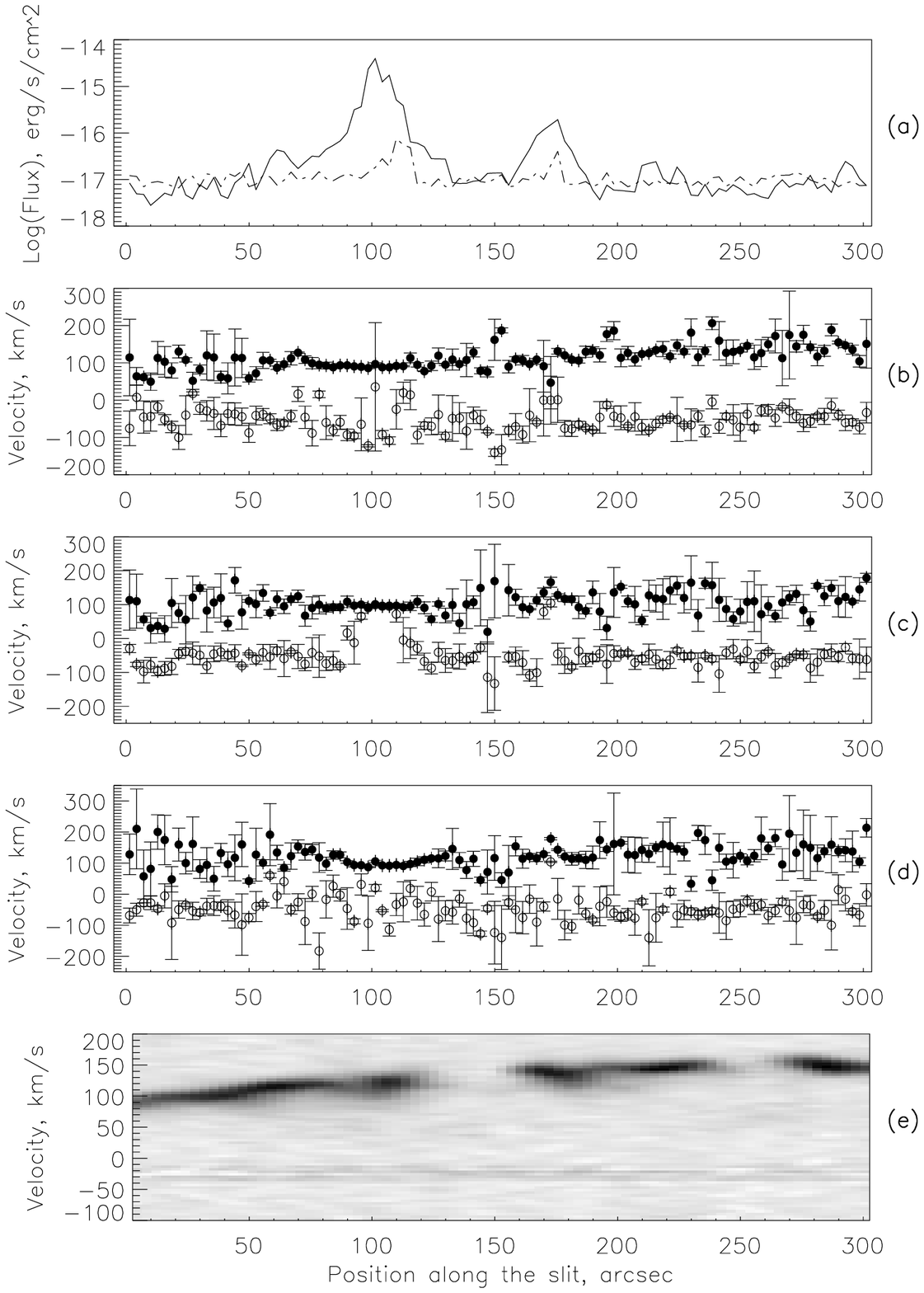}
 
  \caption{The variations of spectral characteristics along the slit PA 180, passing
through the cavity no.\,107. The plots (a)--(e) are similar to
those shown in Fig.\,7.}
\end{figure*}

\section{DISCUSSION}
\subsection{Extraplanar Diffuse Ionized Gas}

The increased ratio of the flux in the forbidden lines of
nitrogen and sulfur to the flux in H$_\alpha$ that we found in the
cavities of NGC~6946 are well known in the extraplanar diffuse
ionized gas (EDIG), found outside of the  planes of galaxies,
visible edge-on, as well as in the EDIG of our Galaxy. Our data
(Figs.~3 and 7--9) show that the entire region of NGC~6946 reveals
gas with a nearly constant velocity of about \mbox{$-40$~km/s},
which is certainly the diffuse ionized gas (EDIG) of our Galaxy,
located outside its plane.

The intensity ratios of the gas along our slits are demonstrated
in Fig.~3. For the slit, passing through the cavity no.\,85, these
ratios  point rather to the photoionization, and for both slits
passing through the cavity no.\,107, and especially in the slit
position PA~147, they are typical for shock excitation. It can be
assumed that these differences reflect the angular sizes of the
DIG/WIM clouds of the Galaxy, and the differences in the
ionization parameters of various clouds.

Note that over the entire length of each slit the ratios of these
lines vary almost chaotically, more precisely, they are constant
within the errors. This bears no resemblance with the intensity
ratio variations of the component coming from the gas of NGC~6946,
which are closely linked with the morphology of the gas and dust
disc of the galaxy.

Note that some figures in Boomsma's thesis  \cite{ref_1:Efremov_n_en}, for
example, on p.~68 and 69, reveal elongated HI clouds of the
foreground with sizes of the order of ten minutes and velocities
ranging from $-67$~to \mbox{$-8$ km/s.} The clouds with similar
characteristics can be as well discerned  in the lane of constant
velocity, present in the figure demonstrating the rotation curve
of NGC~6946 in Boomsma et al. \cite{ref_2:Efremov_n_en},  which the authors do
not even mention. This HI lane with the radial velocity of about
\mbox{$-35 \pm 5$~km/s} exactly corresponds  to the radial
velocity component we have found from the lines of ionized gas; it
extends along the entire length ($26'$) of the rotation curve of
NGC~6946, built along the major axis of the galaxy in this paper,
but the authors \cite{ref_2:Efremov_n_en} do not comment on its presence. We
may suspect that the high-latitude neutral gas of our Galaxy in
the direction of NGC~6946 is concentrated in a layer of clouds
closely adjacent to each other with a small velocity dispersion.
Based on the figures  \mbox{from \cite{ref_1:Efremov_n_en}} we can estimate the
angular sizes of some of these clouds.  They are close to the
characteristic sizes of not completely random (but small,
\mbox{Figs.\,7--9}) radial velocity variations of the ionized gas
of the foreground on our spectrograms. This probably indicates
that the neutral and ionized gas of the Galaxy  at high latitudes
is concentrated in the same clouds, and then we can not consider
it quite so diffuse.

The gas velocity of the foreground we found in a projection to
NGC~6946 is within the velocity of the EDIG of the Galaxy,
observed in this region by Haffner et al.~\cite{ref_14:Efremov_n_en}. This team
and some other authors demonstrate  that this gas of the Galaxy is
characterized by high values of the [NII]/H$_\alpha$ and
[SII]/H$_\alpha$ line intensity ratios, similar to those observed
in the diffuse ionized gas at high (about \mbox{1--2\,kpc})
Z-coordinates in the galaxies, seen edge-on. This way, it was
found in \cite{ref_13:Efremov_n_en} that the [SII]/H$_\alpha$ ratio in NGC~4302
increases from 0.2 at Z $=$ 0\,kpc to 0.6 at Z$=1-2$~kpc.

\subsection{The Origin of Cavities in the Disc of NGC~6946}

Thus, HII regions, located directly at the boundaries of HI
cavities in NGC~6946 show the usual line ratio, typical of
ionization by hot stars.
However, the formation of regions, embordering the cavities, and,
above all, the HII regions and stars, forming an arc along the
southern boundary of the giant (with the diameter of 1.8\,kpc)
cavity no.\,107 is almost evidently related to the sweep-up of the
gas by the expanding shell of the cavity. The absence of spectral
features of the shock wave  in these HII regions means that the
expansion has stopped (or its rate has become slower than the
velocity dispersion in the surrounding gaseous disc). The line
flux ratio  is probably coming back to that, typical of
photoionization within the time scale, shorter than the lifetime
of O-type stars (i.e. less than about 2~Myr). As mentioned above,
there are no indisputable signs of the expansion of
cavities/shells in the radial velocities along the cavities and
their boundaries neither.

The halt in the cavity expansion is not surprising if it was
caused by pressure from within, since the sizes of almost all
cavities are undoubtedly larger than the (effective) thickness of
the gaseous disc of spiral galaxies (in NGC~6946 in particular).
They are hence open to the issue of compressed gas. One way or
another, the cavity no.\,107, exceptional in its size and regular
round shape, reveals no signs of star clusters, which could be a
source of  cavity expansion. Some deviations of radial velocities
inside the hole no. 107 (from the values outside of it) can be
seen in Fig.~9,  but they are oppositely directed for different
gas species, and therefore might hardly be identified as galactic
fountains (gas outflows from the hole).

As emphasized by Boomsma et al.~\cite{ref_2:Efremov_n_en} and many other
authors, it is unclear how the giant sizes of cavities can be
reconciled with the popular hypothesis about the origin of the
cavities under the influence of a pressure source  located inside
of them (from supernovae and/or the stellar wind of O-type stars).
The cavities with effective sizes exceeding the thickness of the
gaseous disc are known in our Galaxy as well. This way, the height
of the gas filaments, limiting by the Z-coordinate the gas,
flowing from the HI supershell in GSH 242-03+37, one of the
largest in our Galaxy (its radius is around 560~pc) is about
1.6\,kpc above and below the galactic \mbox{plane \cite{ref_7:Efremov_n_en}.}
Note, however, that the thickness of the gaseous disc in irregular
galaxies is much  larger than that in spirals, which probably
explains the expansion of gas shells often observed in them (e.g.,
in IC~1613, see~\cite{ref_12:Efremov_n_en}).

Our Figs.~10 and 11 show that  the line intensity ratios inside
the cavities correspond to the shock excitation. The question
arises whether we see the gas in NGC~6946 projected on the cavity
at large \linebreak Z-coordinates, where it has typical for the
EDIG line ratios, similar to those observed at the shock
excitation, or whether the gas of the galactic disc was excited in
the cavities at their formation.

The first hypothesis seems more probable. Inside the cavities of
the gas and dust disc in NGC~6946 the density of gas (and ionized
gas too) is sharply reduced, and at large Z-coordinates it is
obviously equal both over the cavities and beyond. As follows from
Fig.~11, the ratio [SII]/H$_\alpha$ anticorrelates with the flux
in the H$_\alpha$ line, the value of which indicates that inside
the cavities,  the contribution of the disc gas (with ``normal''
photoionization by hot stars) in the integral gas characteristics
along the line of sight is several times smaller. At larger
Z-coordinates in NGC~6946, just like in our and other galaxies, we
should as well find the EDIG with the ionization parameters close
to those, typical of shock excitation. In the projection on the
disc cavities, the contribution of the EDIG at high Z-coordinates
dominates over a small contribution in the cavities of the gaseous
disc. This is why we observe in the cavities (according to our
hypothesis, above them) the ionization parameters, corresponding
to the shock excitation. This hypothesis is also supported by the
argument that the ionization of such nature is preserved in all
holes, including the smallest. Note that there is no correlation
between the value of this ratio and the cavity size. This can be
viewed as an indication that the anomalous ionization in the
region of cavities is not physically related with them.
Apparently, owing to the large amount of cavities in the disc of
NGC~6946, we can for the first time study in the galaxy visible
almost face-on the characteristics of gas high above its plane.

\begin{figure*}

 \includegraphics[width=16cm]{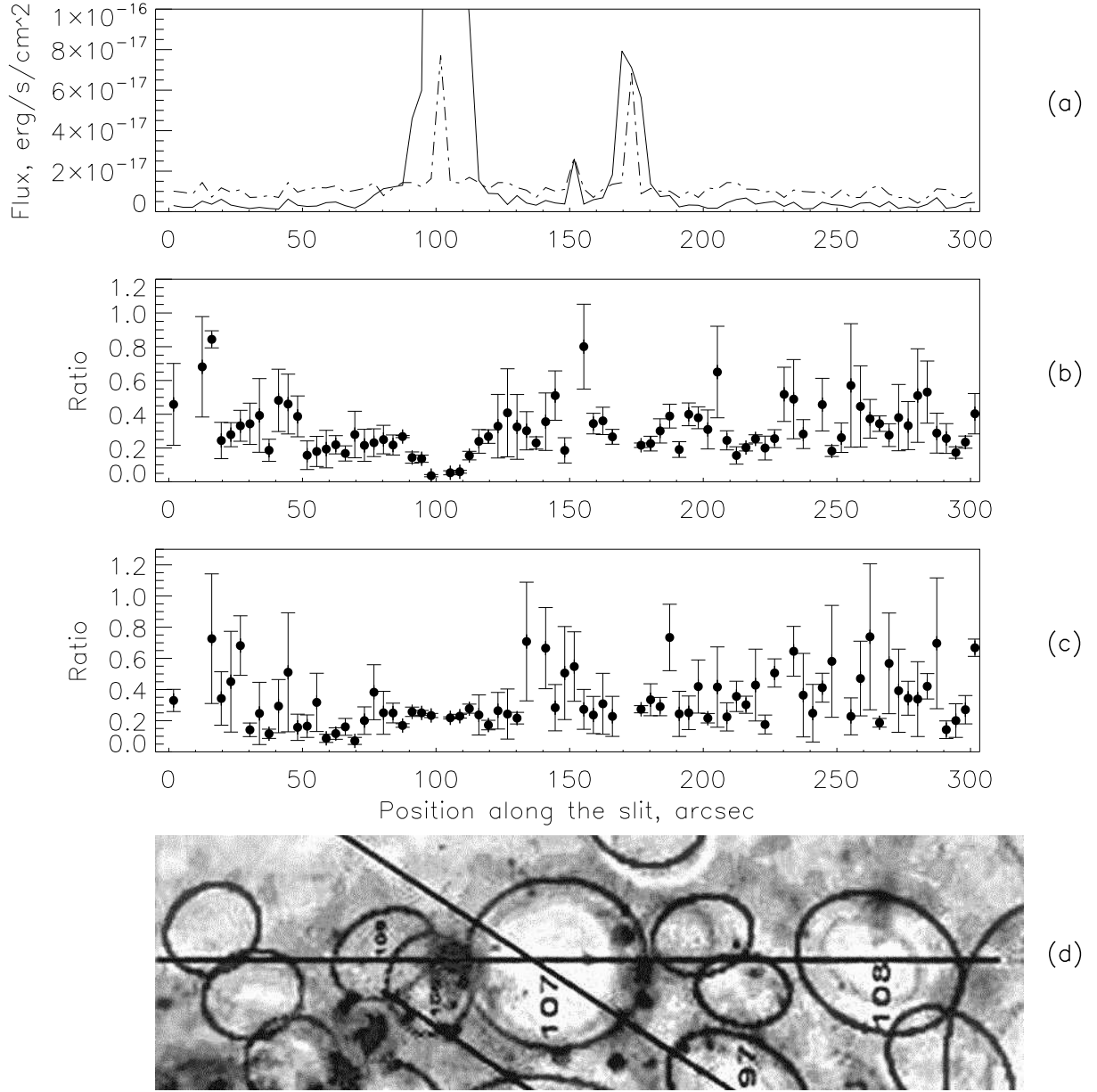}
 \caption{The variations of spectral characteristics along the slit PA 180, passing
through the cavity no.\,107. (a) the fluxes in the  [NII]
$\lambda$6583 line from NGC~6946 (the solid line) and the WIM of
the Galaxy (the dash-dotted line); (b) the flux ratio in the
\mbox{[SII] $\lambda$6717/H$_\alpha$} lines; (c) the flux ratio in
the [NII] $\lambda$6583/H$_\alpha$ lines; (d) the slit position in
the image of the galaxy in the line of HI with respect to the HI
cavities. The HII regions are outlined in black.}
\end{figure*}

\begin{figure*}
 \includegraphics[width=14cm]{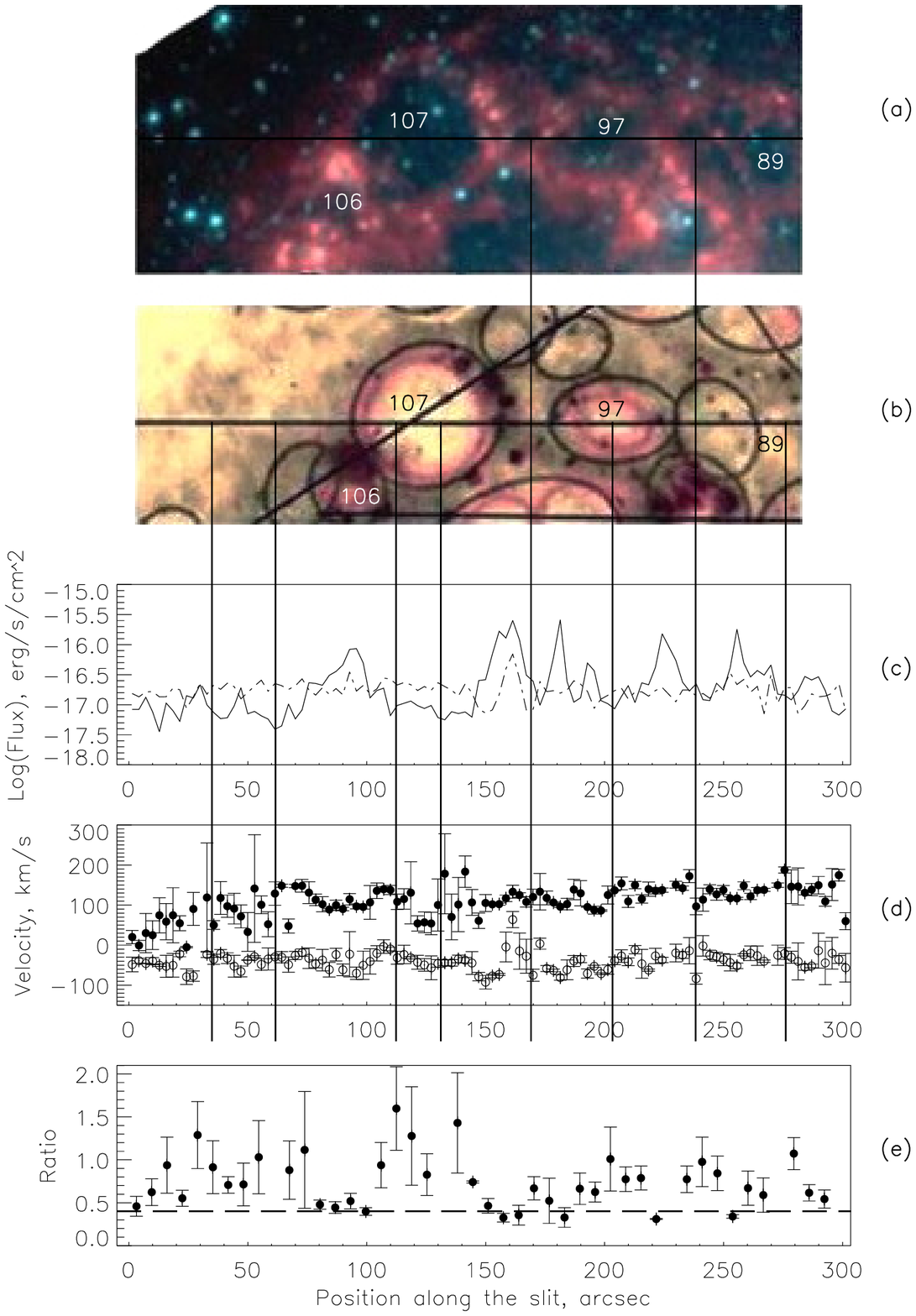}
  \caption{The variations of spectral characteristics along the slit PA 147, passing
through the cavity no.\,107: (a) the image of the galaxy based on
the data obtained by the Spitzer telescope in the far IR region
(see {\tt
http://www.spitzer.caltech.edu/images/2078-sig08-008-The-Fireworks-Galaxy-NGC-6946});
(b) the positions of HI cavities, found in \cite{ref_9:Efremov_n_en} in the
image of the region in HI 21\,cm. The numbers of several regions
are marked; (c) the fluxes in H$_\alpha$ from NGC~6946 (the solid
line) and the WIM of the Galaxy (the dash-dotted line);  (d)
radial velocities of ionized gas of the galaxy NGC~6946 (the upper
curve) and the WIM of the Galaxy (the bottom curve) in the
H$_\alpha$; (e) flux ratio in the  [SII]6717+6731/H$_\alpha$.
lines. Vertical lines indicate the features of  the gas density,
the  line ratios and/or the radial velocity dispersion, the two
longest lines end at the top in the mini-cavities of warm dust,
almost invisible in HI due to the inferior resolution at~21\,cm,
but tangible as the local H$_\alpha$  flux minima   and as local
the maxima of the ratio  [SII]/H$_\alpha$. }
\end{figure*}

Note that there are no star clusters, supernova remnants or X-ray
sources inside the classical cavity no.\,107, surrounded by a
strongly marked gas and dust shell. An arc of HII regions is
located along its southern boundary. Judging from the normal
ionization parameters in them, the respective young stellar groups
were not formed directly under the effect of pressure of the
expanding gas shell, but rather upon reaching a sufficient density
of the swept up gas. However, there arises the question of how
long the spectral features of the shock ionization are maintained.

However, stellar groups are present inside several cavities in
NGC~6946. This way,   inside (but not in the center) of cavity
no.\,85 there is a strange star complex with a diameter of about
600~pc, bounded on the west by a regular semicircle, which is
reigned by a supermassive young cluster,  by contrast not located
in its center  \cite{ref_3:Efremov_n_en}. We hypothesized that this peculiar
complex with its supercluster and the asymmetric cavity (Fig.~1),
in which its is immersed, could form as a result of impact of a
dark \linebreak mini-halo \cite{ref_5:Efremov_n_en}.

A seemingly striking correlation between the intensity, radial
velocity and excitation parameters of the ionized gas in various
regions inside and near this complex, discovered in \cite{ref_5:Efremov_n_en},
can now be simply explained by the fact that at the decreasing
emission intensity  from NGC~6946, the EDIG our Galaxy makes a
more and more growing contribution into the line profiles and
intensities from the foreground gas of the Milky Way, which has much lower radial velocity than that
measured in the gas of NGC~6946, the fact we were previously
unaware of. This, obviously, applies to several results from
\cite{ref_6:Efremov_n_en}. A comparison of Fig.~5 from this work with the image
of NGC~6946 in the infrared range shows that in this figure the
velocities below 80~km/s are only observed in the mini-cavities
visible in the warm dust.

The measurement of radial velocities from the profiles, obtained
with a low spectral resolution, without separation into
components, coming from the Milky Way and from NGC~6946 results in
a decrease in radial velocities, and an increase in the relative
intensity of  forbidden emission lines in the regions with low gas
density of NGC~6946. This explains (but perhaps only partly) the
features of the deep dip inside the peculiar complex, noted but
unguessed in our previous \linebreak work \cite{ref_5:Efremov_n_en}.

Since, as shown in \cite{ref_6:Efremov_n_en}, the peculiar complex as well
contains the expanding gas shells, a wall of one of which is
moving in our direction (detected as early as in 2002
\cite{ref_4:Efremov_n_en}), it is likely that the deep dip \cite{ref_5:Efremov_n_en} is
real. However, its parameters and, above all, the real depth of
the depression, the maximum approach velocity of its shell (or the
rate of one-way gas outflow from it), have yet to be evaluated
anew, taking into account the intensities and velocities of the
ionized extraplanar gas of the Milky Way, the presence of which we
have asserted in this paper.

The sizes of large cavities (of almost all of them in NGC~6946)
exceed the thickness of the gaseous disc in spiral galaxies, which
is a great challenge for the conventional theories of their
formation. As noted by Boomsma et al. \cite{ref_2:Efremov_n_en},
the groups of hot stars which may be responsible for the formation
of cavities are observed only in two or three cases in NGC~6946.
The problem has been studied long and intensively (see, e.g.,
\cite{ref_1:Efremov_n_en, ref_9:Efremov_n_en,
ref_16:Efremov_n_en}), but the causes of cavity formation remain
controversial. In any case, the gravitational impact of a
mini-halo of dark matter \cite{ref_18:Efremov_n_en} is the most
probable cause of formation of the cavity no.\,85, and the
peculiar stellar complex, located on its edge
\cite{ref_5:Efremov_n_en}.

\section{CONCLUSIONS}
\label{sec:summ:Efremov_n_en}

We therefore conclude that  the cavities in the gas and dust disc
of NGC~6946 allow us to observe the EDIG/WIM of our Galaxy---the
extraplanar gas in the projection on the plane of the galaxy.
Obviously, we can only notice the presence of this gas projected
on such regions of the disc of NGC~6946, where the density
(emission intensity) of gas  with   ``normal'' line ratios is
small,  i.e. projected on the disc cavities.
We are not yet aware whether such an ``anomaly'' of the line
intensity ratio was ever observed inside the cavities (more
precisely, as we assume, projected on the cavity) in other
galaxies. To verify our findings, we plan similar observations of
several nearby spiral galaxies.

Across the disc of NGC~6946, we found two components in the radial
velocities of lines of ionized gas---besides the component with
varying intensity and velocity, ionized gas with almost constant
radial velocity on the average amounting to $-40$~km/s is present
throughout the galaxy. The flux of the latter component varies
along the slit only slightly and randomly, just like the line
ratios, and its spectral characteristics are close to those,
observed in the warm diffuse gas high above the Galactic plane.
There is no doubt that this component originates in the Milky Way
at high Z-coordinates.

What we observe in the projection on the cavities of the gas and
dust disc of NGC~6946, is the EDIG of this galaxy, located high
above its plane. That explains the typical EDIG/WIM ratio of the
 [SII]/H$_\alpha$ lines, exceeding 0.4, which is observed in all without exception, cavities of various
sizes (including such minuscule holes, observed only in the warm
dust in the images obtained with the IR Spitzer Space Telescope).

The formation of cavities (bubbles, holes, shells, supershells) in
the galactic discs of neutral hydrogen is usually attributed to
the fallout  of fast gas clouds on the disc, or to the effect of
the supernova explosion energy and stellar winds from hot stars on
the surrounding gas. The second hypothesis is however confirmed
only in few cases, and does not pass as a general rule. In
particular, the arguments against it include the usual lack of
appropriate clusters inside the cavities, and the fact that no cavities
are revealed around a great number of star clusters of suitable
age and luminosity \cite{ref_15:Efremov_n_en}.

We have to finish our paper stating that we have found the need to
exclude the emission of extraplanar gas of our Galaxy in the study
of NGC~6946, but found no signs of expanding cavities/shells in
the gaseous disc of the galaxy, nor the evidence of collision
ionization in the HII regions, bordering the cavities. Somewhat
surprisingly, we saw these evidences inside the cavities, and most
probably above them. A double conspiracy of spectral lines
of ionized gas
in NGC~6946, showing the signs of shock excitation not in the
regions expected   is a subject of further investigation.

The study is based on the observational data obtained at the 6-m
BTA telescope funded by The Ministry of Education and Science
of the Russian Federation (registration no. 01-43).

\begin{acknowledgements}

We are very grateful to A.~V.~Moiseev for a detailed critical
feedback on the manuscript, which led to its considerable
improvement, especially in the plots used. O.~V.~Egorov is
grateful for the financial support of the non-profit Dmitry
Zimin's Dynasty Foundation and the Russian Foundation for Basic
Research (grant no. 10-02-00091) (directed by T.~A.~Lozinskaya);
Yu.~N.~Efremov thanks the Russian Foundation for Basic Research
(grant no. 10-02-00178) (directed by A.~D.~Chernin). We
particularly thank R.~Boomsma, who has built the profiles of
radial velocities of neutral hydrogen along our slits based on his
observations with the WSRT.
\end{acknowledgements}

\quad

\flushright{\textit{Translated by A.~Zyazeva}}


\begin{thebibliography}{99}


\bibitem{ref_1:Efremov_n_en}
 R.~Boomsma,  Thesis, {\footnotesize{\tt http://
dissertations. \linebreak
ub.rug.nl/faculties/science/2007/r.boomsma/}}

\bibitem{ref_2:Efremov_n_en}
R.~Boomsma et~al.,  A\&A~{\bf 490}, 555 (2008).

\bibitem{ref_9:Efremov_n_en}
I.~Bagetakos, E.~Brinks, F.~Walter, et al., AJ~{\bf 141}, ..23  (2011).

\bibitem{ref_16:Efremov_n_en}
D.~R.~Weisz, E.~V.~Skillman,  J.~M.~Cannon, et
al., ApJ~{\bf 704}, 1538 (2009).

\bibitem{ref_17:Efremov_n_en}
G.~Tenorio-Tagle and   P.~Bodenheimer,
Ann. Rev. A\&A {\bf 26}, 145 (1988). 

\bibitem{ref_15:Efremov_n_en}
Yu.~N.~Efremov,  A\&A Trans.~{\bf 21}, 251 (2002). 

\bibitem{ref_18:Efremov_n_en}
 K.~Bekki and  M.~Chiba,  ApJ~\textbf{637}, (2006).


\bibitem{ref_3:Efremov_n_en}
S.~Larsen, Yu.~N.~Efremov,  B.~G. Elmegreen, et
al., ApJ~{\bf 567}, 896 (2002).

\bibitem{ref_4:Efremov_n_en}
 Yu.~N.~Efremov,  S.~A.~Pustilnik, A.~Y.~Kniazev,
et al.,  A\&A~{\bf 389}, 855 (2002).

\bibitem{ref_5:Efremov_n_en}
 Yu.~N.~Efremov,  V.~L.~Afanasiev, E.~J.~Alfaro,
et al.,   MNRAS~{\bf 382}, 481 (2007).

\bibitem{ref_6:Efremov_n_en}
 M.~Carmen S\'anchez Gil, E.~J. Alfaro, and
E.~P\'erez,  ApJ~{\bf }702, 141 (2009).

\bibitem{ref_7:Efremov_n_en}
 N.~M.~McClure-Griffiths, A.~Ford,  D.~J.~Pisano,
et al.,  ApJ~{\bf 638}, 196 (2006).

\bibitem{ref_8:Efremov_n_en}
 V.~L.~Afanasiev and A.~V.~Moiseev, Astron. Lett. \textbf{31}, 194 (2005).


\bibitem{ref_19:Efremov_n_en}
F.~Bonnarel et al.,  A\&A~{\bf 189}, 59
(1988).


\bibitem{ref_10:Efremov_n_en}
 A.~D.~Chernin, Yu.~N.~Efremov, and
P.~A.~Voinovich, MNRAS~{\bf 275},  313 (1995).



\bibitem{ref_11:Efremov_n_en}
 A.~Moiseev, I.~Karachentsev, and S.~Kaisin,
MNRAS~{\bf 403}, 1849  (2010).

\bibitem{ref_14:Efremov_n_en} L.~M.~Haffner, R.~J.~Reynolds,  S.~L.~Tufte, et
al., 
ApJS~{\bf 149}, 405 (2003).

\bibitem{ref_13:Efremov_n_en}
 J.~A.~Collins and  R.~J.~Rand, ApJ~{\bf 551}, 57
(2001).


\bibitem{ref_12:Efremov_n_en}
 S.~Silich, T.~Lozinskaya,  A.~Moiseev, et al.,
A\&A~{\bf 448}, 123  (2006).












\end{thebibliography}
\end{document}